\documentclass{article}

\PassOptionsToPackage{numbers, compress}{natbib}

\usepackage[preprint]{neurips_2026}
\usepackage[utf8]{inputenc}   
\usepackage[T1]{fontenc}      
\usepackage{hyperref}         
\usepackage{url}              
\usepackage{booktabs}         
\usepackage{amsmath, amssymb, amsfonts}
\usepackage{nicefrac}         
\usepackage{microtype}        
\usepackage{xcolor}           
\usepackage{graphicx}
\usepackage{float}
\usepackage{multirow}
\usepackage{array}
\usepackage{longtable}
\usepackage{makecell}
\usepackage{caption}
\usepackage{subcaption}
\usepackage{enumitem}
\usepackage{pgfplots}
\pgfplotsset{compat=1.18}

\hypersetup{colorlinks=true, linkcolor=blue, urlcolor=cyan, citecolor=blue}

\title{SraVaani 1.0: Scaling Inclusive Speech Recognition\\ for Indic Languages}

\author{%
  Sujith Pulikodan \quad Agneedh Basu \quad Pavan Kumar J \quad Pranav D Bhat \\
  \textbf{Suryansh Shukla} \quad \textbf{Nihar Desai} \quad \textbf{Prasanta Kumar Ghosh} \\[0.6em]
  SPIRE Lab, Indian Institute of Science (IISc), Bangalore \\
  ARTPARK@IISc, Indian Institute of Science (IISc), Bangalore \\
  \texttt{sujith@artpark.in} \\
}

\begin{document}

\maketitle

\begin{abstract}
  India's linguistic landscape spans over 700 languages and thousands of dialects, 
  yet the vast majority of automatic speech recognition (ASR) systems support only a small 
  fraction of this diversity. We present \textbf{SraVaani-1.0}, a multilingual ASR model 
  covering \textbf{65 Indian languages and dialects}, many of which currently have no publicly
   available or competing ASR system. SraVaani-1.0 is built on a FastConformer architecture and
    trained from scratch through a three-stage pipeline. In the first stage, we perform self-supervised 
    pretraining on \textbf{31,255 hours} of unlabelled speech from the VAANI corpus using a contrastive 
    learning objective. In the second stage, we introduce an \textit{audio--image representation alignment} 
    stage that leverages the paired images and speech available in the VAANI corpus; this multimodal
     alignment encourages the speech encoder to learn semantically richer representations 
     by exploiting the relationship between visual context and spoken content, thereby 
     improving downstream recognition, particularly for low-resource languages. In the 
     final stage, the aligned encoder is fine-tuned end-to-end using a
     Hybrid Token-and-Duration Transducer (TDT)--CTC decoder 
      on \textbf{31,263 hours} of labelled multilingual Indian speech compiled from \textbf{24 public datasets}
       spanning 65 languages and dialects. We evaluate SraVaani-1.0 against three state-of-the-art multilingual
        ASR systems---IndicConformer-600M-Multilingual, Sarvam Saaras v3, 
        and Gemini 3 Flash---across eight benchmarks: \textbf{Common Voice}, 
        \textbf{FLEURS}, \textbf{IndicTTS}, \textbf{Kathbath}, \textbf{RESPIN}, \textbf{GramVaani},
         \textbf{MUCS}, and \textbf{VAANI}. SraVaani-1.0 achieves the lowest word error rate (WER) 
         on a large number of language--dataset pairs while remaining competitive with the best-performing 
         systems on high-resource languages. Most importantly, it is the only opensource evaluated model that provides 
         transcription capability for \textbf{multiple low-resource and tribal Indian languages}, which are
          assessed exclusively on the VAANI benchmark. These results demonstrate that large-scale 
          self-supervised learning, multimodal representation alignment, and multilingual 
          supervised fine-tuning together enable robust ASR for a significantly broader
           spectrum of India's linguistic diversity than previously possible.
\end{abstract}

\section{Introduction}
\label{sec:introduction}

India is home to one of the most linguistically diverse populations on Earth.
The Eighth Schedule of the Indian Constitution recognises 22 official languages,
yet the country harbours over 700 living languages and thousands of dialects
spanning four major language families: Indo-Aryan, Dravidian, Tibeto-Burman,
and Austro-Asiatic. Despite this richness, the vast majority of automatic speech recognition (ASR)
research has focused on a handful of high-resource languages,
leaving most Indic languages severely underserved by
existing technology. The practical consequences are significant.
Government services, healthcare, education, and accessibility tools increasingly
rely on speech interfaces, yet rural and tribal communities that speak
non-scheduled languages are excluded from these benefits.
Building ASR systems that generalise across India's linguistic landscape
therefore constitutes both a technical challenge and a matter of digital equity.

Recent large-scale multilingual ASR models such as
Whisper~\cite{radford2023whisper},
Google USM~\cite{zhang2023google}, and
XLS-R~\cite{babu2022xls}
have expanded language coverage substantially, but still provide limited or no
support for the majority of Indic languages---particularly low-resource tribal
and dialect variants.
Specialised Indic systems achieve strong performance on the scheduled languages
they target---\textbf{IndicConformer-600M-Multilingual}\cite{ai4bharat_indicconformer_2025} covers the 22 Eighth
Schedule languages, and \textbf{Sarvam Saaras v3}~\cite{sarvam2024saaras} covers also covers 22, but offer no coverage beyond that set.
No existing system provides transcription capability across the full breadth of
India's linguistic diversity.

In this report, we present \textbf{SraVaani-1.0}\footnote{\url{https://huggingface.co/ARTPARK-IISc/SraVaani-1.0}}, a multilingual automatic speech recognition (ASR) model covering a subset of Indian languages that have limited representation in existing ASR systems.
SraVaani-1.0 is trained through a three-stage pipeline that combines
self-supervised learning, multimodal representation learning, and supervised
fine-tuning.
In the first stage, we perform self-supervised pretraining of a FastConformer
encoder\cite{rekesh2023fast}  on \textbf{31,255 hours} of unlabelled spontaneous speech from the
\textbf{VAANI} corpus~\cite{vaani2024}, collected across 28 states and union
territories of India covering 105 languages, using contrastive learning
objective~\cite{baevski2020wav2vec} (Section~\ref{sec:pretraining}).
In the second stage, we further adapt the pretrained encoder using the
\textbf{11.85 million} paired audio--image samples that VAANI's picture-prompt
collection protocol yields at no additional annotation cost, through an
audio--image representation alignment objective
(Section~\ref{sec:alignment}).
This multimodal training encourages the encoder to learn semantically richer
acoustic representations by exploiting the correspondence between spoken
utterances and their associated visual context, thereby improving recognition
performance, particularly for low-resource languages.
Finally, the aligned encoder is fine-tuned end-to-end on \textbf{ 31,263  hours}
of transcribed multilingual Indian speech compiled from \textbf{24 public
datasets} spanning \textbf{65 languages and dialects}
(Section~\ref{sec:finetuning}).
We employ a Hybrid Token-and-Duration Transducer (TDT)--CTC decoder, jointly
optimising the transducer and CTC objectives to produce a single multilingual
ASR model capable of recognising a broad spectrum of Indian languages and
dialects. The resulting model supports \textbf{65 Indian languages and dialects}, many of
which currently have no publicly available ASR system, making SraVaani-1.0 the
broadest-coverage multilingual Indic ASR model reported to date.

Our main contributions are threefold.
First, we propose a three-stage training framework that combines
self-supervised pretraining, audio--image representation alignment, and
supervised multilingual fine-tuning to improve ASR performance, particularly
for low-resource Indian languages.
Second, we develop a FastConformer multilingual ASR model trained
entirely on publicly available speech datasets, providing transcription
capability for \textbf{65 Indian languages and dialects}, including many
low-resource and tribal languages for which no previous ASR system exists.
Third, we present a comprehensive evaluation across eight public
benchmarks---CommonVoice\cite{ardila2020common}, FLEURS\cite{conneau2023fleurs}, IndicTTS\cite{indictts2023}, Kathbath\cite{javed2023indicsuperb}, RESPIN\cite{NEURIPS2025_d88714d4}, GramVaani\cite{bhanushali2022gram},
MUCS\cite{diwan2021multilingual}, and VAANI---comparing SraVaani-1.0 on accuracy against  Sarvam Saaras v3, IndicConformer-600M-Multilingual
and Gemini~3 Flash~\cite{google2024gemini} (Section~\ref{sec:exp_main}).

\section{Model Architecture}
\label{sec:model}

\subsection{Overview}

SraVaani-1.0 is trained using a three-stage pipeline. First, the encoder is pretrained in a self-supervised manner using a contrastive learning objective on large-scale unlabelled speech to learn robust acoustic representations. Second, the pretrained encoder is further adapted through an audio--image representation alignment stage, where speech representations are aligned with image embeddings extracted from a frozen vision encoder using the picture--prompt pairs provided by the VAANI dataset (Section~\ref{sec:alignment}). This stage does not require any speech transcriptions and aims to inject semantic information into the speech encoder. Finally, the aligned encoder is fine-tuned end-to-end on labelled multilingual Indian speech using a Hybrid TDT-CTC decoder. All stages are implemented using the \textbf{NVIDIA NeMo} framework~\cite{nemo2019}. A high-level summary of the three training stages is presented in Table~\ref{tab:summary} (Section~\ref{sec:finetuning}).
\subsection{Encoder and Audio Front-end}
\label{sec:encoder}

The backbone acoustic encoder used throughout all three training stages is the \textbf{FastConformer} architecture. FastConformer is an efficient variant of the Conformer architecture~\cite{gulati2020conformer} that replaces the standard convolutional subsampling module with a depthwise-strided convolutional subsampling scheme, reducing the input sequence length by a factor of $8\times$ before the self-attention layers. This substantially lowers the computational cost of the attention mechanism while preserving modelling capacity, enabling efficient training and inference on long speech sequences. The encoder comprises 17 Transformer layers with a model dimension of 1,024 and 8 attention heads. Each layer employs a feed-forward network with a $4\times$ expansion factor and a convolution module with a kernel size of 9. Relative positional encoding is used to model long-range temporal dependencies, while both the dropout and attention dropout rates are set to 0.1.

All audio is resampled to 16~kHz and converted into 128-dimensional log-Mel filterbank features extracted using a 25~ms Hann window with a 10~ms frame shift and a 512-point FFT. The features are normalized using per-feature mean and variance normalization, and a dithering factor of $10^{-5}$ is applied during feature extraction to improve numerical stability. To improve model generalization, SpecAugment~\cite{park2019specaugment} is applied consistently during both the self-supervised pretraining and supervised fine-tuning stages. The augmentation policy consists of two frequency masks with a maximum width of 27 Mel bins and ten time masks, each spanning at most 5\% of the utterance duration.
\section{Self-Supervised Pretraining}
\label{sec:pretraining}

\subsection{Pretraining Data: VAANI}
\label{sec:pretrain_data}

Pretraining is performed on the VAANI corpus , a large-scale collection of spontaneous speech gathered across 28 states and 3 union territories and spanning 165 districts and 105 languages, which gives the model broad exposure to regional accent, dialect, and acoustic-environment variation. No transcriptions are used; the corpus is consumed purely as unlabeled audio for self-supervised representation learning. After segmentation and filtering to a [0.5, 25.0] s utterance-length window, the corpus comprises 21,087,852 utterances totalling 29,912 hours of speech, with a mean utterance duration of 5.1 s and a median of 4.3 s. This is partitioned 95/5 into a training split of 20,033,459 utterances (28,418 h) and a validation split of 1,054,393 utterances (1,494 h). The split is stratified across all 165 districts. 
\subsubsection{Language coverage}

The VAANI dataset covers 105 Indian languages and dialects, including Hindi, Bengali, Telugu, Kannada, Marathi, Tamil, Odia, Chakma, Bhojpuri, Garo, Maithili, Nepali, Chhattisgarhi, Assamese, Malayalam, English, Gujarati, Nagamese, Punjabi, Manipuri, Mizo, Rajasthani, Urdu, Marwari, Garhwali, Wancho, Magahi, Angika, Karbi, Bajjika, Konkani, Halbi, Kokborok, Tulu, Haryanvi, Sambalpuri, Kashmiri, Khortha, Kumaoni, Sadri, Surjapuri, Khariboli, Surgujia, Nimadi, Malvani, Kurukh, Bundeli, Idu Mishmi, Angami, Sumi, Gondi, Awadhi, Lepcha, Desia, Santali, Bearybashe, Khandeshi, Nyishi, Chakhesang, Ao, Rengma, Sindhi, Rongmei, Bhili, Bagheli, Koya, Sikkimese, Jaipuri, Tangkhul, Tagin, Bhatri, Powari, Malvi, Kurmali, Dorli, Pahadi, Yimchunger, Shekhawati, Sangtam, Bagri, Thethi, Lambani, Sylheti, Liangmai, Wagdi, Zeme, Galo, Duruwa, Thadou, Hajong, Dogri, Mewati, Harauti, Mewari, Vaiphei, Rajbanshi, Limbu, Phom, Sirmauri, Agariya, Baghati, Paniya, Mara, and Kuki. This broad linguistic coverage spans constitutionally recognized languages, regional languages, tribal languages, and local dialects from across India.

\subsection{SSL Objective: Contrastive Loss}

Pre-training uses NeMo's \texttt{SpeechEncDecSelfSupervisedModel} with a
wav2vec~2.0-style~\cite{baevski2020wav2vec} contrastive objective.
Input log-mel spectrograms (128 mel bins) are corrupted by
\texttt{SpectrogramAugmentation} with 2 frequency masks (width 27 bins) and
10 time masks (each spanning 5\% of the utterance).
Targets are constructed directly from the \emph{uncorrupted} spectrogram:
consecutive blocks of 8 frames---matching the encoder's $8\times$
depthwise-striding subsampling, so one target aligns with one encoder output
frame and covers 80\,ms of audio---are flattened to 1{,}024 dimensions and
mapped through a linear projection to $d=256$.
The vector-quantiser is disabled, so targets are continuous projections rather
than discrete codebook entries.
In parallel, a \texttt{ConvASRDecoderReconstruction} head (no stride or
non-stride layers) projects the 1{,}024-dimensional encoder output to the same
256-dimensional space, giving the prediction $\mathbf{z}_t$.

For every masked position $t$, the model must identify its own target
$\mathbf{q}_t$ against $|\mathcal{N}|=50$ distractors drawn from the masked
positions of \emph{all} utterances in the batch:
\[
  \mathcal{L}_\text{SSL} = -\sum_{t \in \mathcal{M}} \log
    \frac{\exp\!\bigl(\mathrm{sim}(\mathbf{z}_t,\,\mathbf{q}_t)/\kappa\bigr)}
         {\exp\!\bigl(\mathrm{sim}(\mathbf{z}_t,\,\mathbf{q}_t)/\kappa\bigr)
          + \displaystyle\sum_{j \in \mathcal{N}}
            \exp\!\bigl(\mathrm{sim}(\mathbf{z}_t,\,\mathbf{q}_j)/\kappa\bigr)}
\]
where $\mathrm{sim}(\cdot,\cdot)$ is cosine similarity, $\kappa=0.1$ is the
logit temperature, and $\mathcal{M}$ is the set of encoder frames whose
constituent spectrogram frames are more than 80\% masked.
The loss is summed over masked positions and scaled by $16/B$ for batch
size~$B$.

\subsection{Pretraining Configuration}

Self-supervised pretraining is performed with NVIDIA NeMo on top of PyTorch
Lightning, using distributed data-parallel training in FP32 precision with
synchronised batch normalisation.
Each GPU processes 128 utterances per step and gradients are accumulated over
16 steps before each update, with gradients clipped to a global norm of 1.0.
The model is optimised with AdamW ($\beta_1 = 0.9$, $\beta_2 = 0.999$, weight
decay $1\times10^{-4}$) under a Noam annealing schedule with a base scaling
factor of $5\times10^{-3}$ and 2{,}000 warmup steps; given
$d_\text{model} = 1{,}024$, the learning rate peaks at $3.49\times10^{-6}$ at
the end of warmup, decays as $t^{-0.5}$, and is floored at a minimum of
$10^{-6}$.
Training runs for 70 epochs over the 20,033,459 filtered
training utterances, with validation evaluated at the end of each epoch and the
three checkpoints achieving the lowest validation loss retained.

\subsection{Training Curves}

\begin{figure}[H]
\centering

\begin{minipage}{0.48\textwidth}
  \centering
  \includegraphics[width=\linewidth]{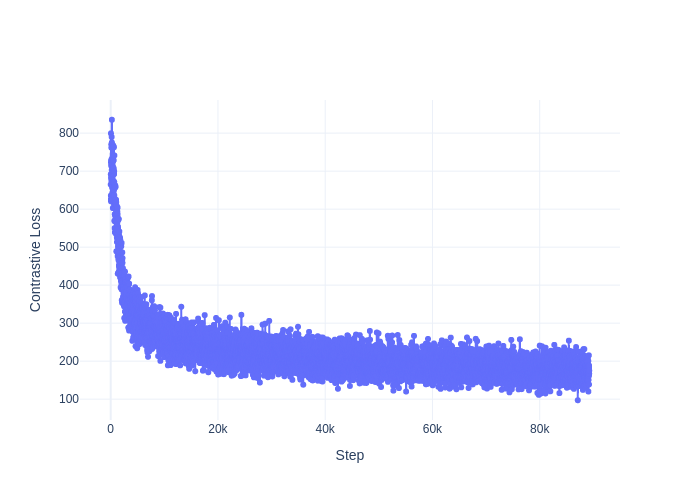}
\end{minipage}
\hfill
\begin{minipage}{0.48\textwidth}
  \centering
  \includegraphics[width=\linewidth]{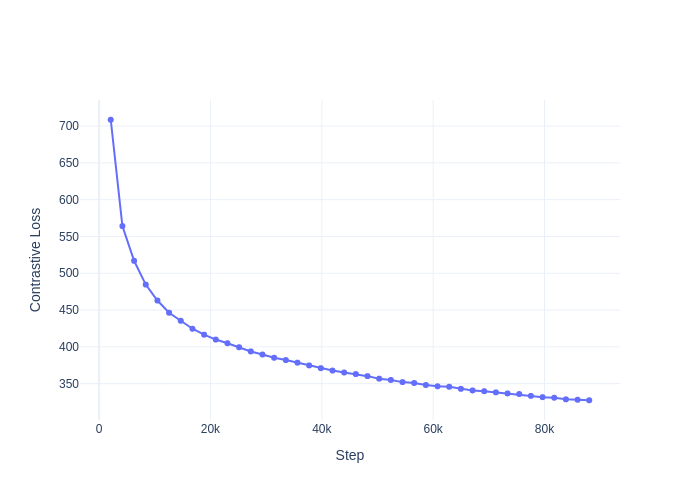}
\end{minipage}

\caption{Training (left) and validation (right) contrastive loss vs.\ step.}
\label{fig:loss_curves}

\end{figure}

Figure~\ref{fig:loss_curves} shows the contrastive loss over training steps
for both the training and validation sets.
The training loss (per-step) drops sharply in the first $\sim$10k steps from
$\approx$820 to below 300, then continues to decrease gradually, reaching
$\approx$150 by 90k steps.
The validation loss (per-epoch checkpoint) follows a smooth monotonic descent
from $\approx$710 to $\approx$330, confirming that the model generalises well
and does not overfit despite its large capacity.

\section{Audio--Image Representation Alignment}
\label{sec:alignment}

\subsection{Motivation}

\begin{figure}[h]
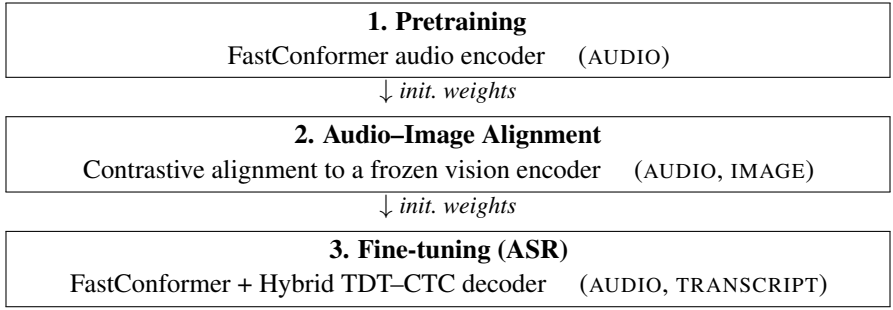

\centering
\begin{tabular}{c}
\fbox{\begin{minipage}{0.82\textwidth}
\centering \textbf{1. Pretraining} \\[0.2em]
FastConformer audio encoder \quad ($\textsc{audio}$)
\end{minipage}} \\[0.35em]
$\downarrow$ {\small\itshape init.\ weights} \\[0.35em]
\fbox{\begin{minipage}{0.82\textwidth}
\centering \textbf{2. Audio--Image Alignment} \\[0.2em]
Contrastive alignment to a frozen vision encoder
\quad ($\textsc{audio}$, $\textsc{image}$)
\end{minipage}} \\[0.35em]
$\downarrow$ {\small\itshape init.\ weights} \\[0.35em]
\fbox{\begin{minipage}{0.82\textwidth}
\centering \textbf{3. Fine-tuning (ASR)} \\[0.2em]
FastConformer + Hybrid TDT--CTC decoder
\quad ($\textsc{audio}$, $\textsc{transcript}$)
\end{minipage}}
\end{tabular}
\caption{Three-stage pipeline: audio pretraining $\rightarrow$ audio--image
alignment $\rightarrow$ ASR fine-tuning, with encoder weights carried forward
between stages.}
\label{fig:three_stage}
\end{figure}

The conventional ASR recipe uses audio-only data during pretraining and
audio--text pairs during fine-tuning, leaving other readily available
multimodal sources unexploited.
For most Indian languages, high-quality transcription is expensive,
time-consuming, and requires language expertise, so the amount of labelled
speech available for fine-tuning remains the binding constraint on accuracy.
We exploit this by inserting an intermediate \textbf{audio--image alignment}
stage between self-supervised pretraining (Section~\ref{sec:pretraining}) and
supervised fine-tuning (Section~\ref{sec:finetuning}).  During this stage the
audio encoder is trained to align its representations with semantic
representations extracted from the prompting images by a frozen, pretrained
vision encoder.  The stage is entirely \emph{transcription-free}: it consumes
no text labels, and it introduces no audio beyond what pretraining has already
seen.  The resulting three-stage pipeline is shown in
Figure~\ref{fig:three_stage}.

\subsection{Approach}

A frozen pretrained image encoder produces the image representations, which are
precomputed once and cached on disk.  The audio encoder is then trained to
align its own representations with these image embeddings; \emph{all} 17
FastConformer blocks are updated during this stage.  To make the two spaces
compatible we introduce an \textbf{alignment head}---an attention-pooling layer
followed by an MLP---that projects the variable-length audio representation
into the image embedding space.  Both the encoder and the alignment head are
optimised; the vision tower is never updated. After alignment, the alignment head is discarded and only the adapted encoder
is carried forward to the ASR stage, where it is coupled with the Hybrid
TDT--CTC decoder and trained on transcribed speech.

\subsubsection{Alignment Objective}

All configurations are trained with the sigmoid (SigLIP-style)
contrastive loss~\cite{zhai2023sigmoid}, using a learnable temperature
$t$ and bias $b$ and in-batch negatives gathered across all GPUs:
\[
  \mathcal{L}_\text{align} =
    -\sum_{i}\sum_{j} \log \sigma\!\bigl(y_{ij}\,(t\,s_{ij} + b)\bigr),
  \qquad
  y_{ij} = \begin{cases} +1 & i = j \\ -1 & i \neq j \end{cases}
\]
where $s_{ij}$ is the similarity between audio $i$ and image $j$.
Unlike softmax-based contrastive losses, the sigmoid form treats every pair
independently and therefore scales cleanly with the number of gathered
negatives.

Two similarity functions are used.  For the single-token configuration,
$s_{ij}$ is the plain cosine similarity between the $L_2$-normalised
projected audio vector $\mathbf{a}_i$ and image vector $\mathbf{v}_j$:
\[
  s_{ij} = \mathbf{a}_i^{\top}\mathbf{v}_j .
\]
For the multi-token configurations we use an asymmetric
\textsc{MaxSim} score in the spirit of late-interaction
retrieval~\cite{khattab2020colbert}: each of the $K_a$ audio queries takes its
best-matching image token, and the results are averaged over queries (padded
image tokens are masked out):
\[
  s_{ij} = \frac{1}{K_a}\sum_{a=1}^{K_a}
           \max_{v \le K_v} \langle \mathbf{q}^{(i)}_a,\; \mathbf{V}^{(j)}_v \rangle .
\]

\subsection{Alignment Configuration}
\label{sec:alignment_config}

Our audio--image alignment model employs the \textbf{SigLIP2-Large} visual encoder (\texttt{patch16-384})~\cite{tschannen2025siglip2}. The FastConformer encoder is initialized from the self-supervised pretrained checkpoint and all 17 encoder blocks are fine-tuned during alignment. The encoder output is compressed into $K_a=16$ audio tokens using a multi-query attention pooling module, followed by a projection MLP that maps the 1,024-dimensional audio representations to the 1,024-dimensional SigLIP embedding space. On the image side, the frozen SigLIP2 encoder produces 576 patch embeddings, from which the top 16 tokens are selected based on their $L_2$ norms and subsequently $L_2$-normalized. Audio--image similarity is computed using the \textsc{MaxSim} operator, which aggregates the maximum similarity between each audio token and the selected image tokens. The model is trained for 200,000 optimization steps using the SigLIP-style sigmoid contrastive loss, while the optimizer, distributed data parallel (DDP) gathering strategy, and other training hyperparameters remain identical to those used throughout the rest of the training pipeline.

\subsection{Alignment Data}
\label{sec:align_data}

The alignment stage uses \textbf{11{,}848{,}593} audio--image pairs, spanning
\textbf{287K} unique images and \textbf{16{,}580.36} hours of audio, all drawn
from the same VAANI partition used for self-supervised pretraining
(Section~\ref{sec:pretrain_data}). This is a deliberate design choice: because no additional or unseen audio is
introduced at this stage, any downstream gain \emph{cannot} be explained by
exposure to new speech data.  It must instead arise from the alignment signal
applied to audio the encoder has already seen.  No overlaping data from any evaluation set
is used at any stage---pretraining, alignment, or fine-tuning.

\section{Supervised Fine-tuning}
\label{sec:finetuning}

\subsection{Decoder: FastConformer-Hybrid-TDT-CTC}

Supervised fine-tuning uses NeMo's hybrid transducer--CTC architecture
(\texttt{EncDecHybridRNNTCTCBPEModel}).
The SSL-pretrained FastConformer encoder (17 layers, $d_\text{model}=1{,}024$,
8 attention heads, $8\times$ depthwise-striding subsampling) is retained and two
decoder heads are attached and trained jointly.
The primary head is a Token-and-Duration Transducer~\cite{xu2023efficient} with
a single-layer RNN prediction network of hidden size 640 and a joint network of
hidden size 640 with ReLU activation, computed with a fused joint batch of 4 to
bound activation memory; the duration vocabulary is $\{0,1,2,3,4\}$ with
$\sigma=0.02$ and $\omega=0.1$.
The auxiliary head is a linear CTC decoder over the same encoder output.
The two are combined as
\[
  \mathcal{L} = (1-\lambda_\text{CTC})\,\mathcal{L}_\text{TDT}
              + \lambda_\text{CTC}\,\mathcal{L}_\text{CTC},
  \quad \lambda_\text{CTC} = 0.3,
\]
and the self-supervised contrastive is disabled throughout.
Inference uses batched greedy TDT decoding.

\subsection{Tokeniser}

A SentencePiece BPE tokeniser with 5{,}000 subword units, trained on the
transcription text of the combined multilingual corpus, provides a single shared
vocabulary spanning all Indian scripts represented in the data.

\subsection{Initialisation}

Fine-tuning warm-starts from the  checkpoint of a prior audio image align stage.
Optimiser and epoch state are not carried over: the run begins from those
weights with a fresh schedule, and trains at a low learning rate of $10^{-5}$ so
as to preserve the acoustic representations acquired during pretraining.

\subsection{Fine-tuning Configuration}

Fine-tuning is performed using NVIDIA NeMo on PyTorch Lightning with distributed data-parallel (DDP) training in \texttt{bf16-mixed} precision and synchronized batch normalization. Each GPU processes a batch of 48 utterances without gradient accumulation, and gradients are clipped to a global norm of 1.0. Under this configuration, each training epoch consists of 92{,}845 optimization steps.

The model is optimized using AdamW ($\beta_1 = 0.9$, $\beta_2 = 0.98$, weight decay $= 10^{-3}$) with a Noam learning rate scheduler configured with 10{,}000 warmup steps and a minimum learning rate of $10^{-5}$. Since the base learning rate is also set to $10^{-5}$, the learning rate remains constant at $10^{-5}$ throughout training after the warmup phase. Training is conducted for up to 50 epochs, with validation word error rate (WER) evaluated after each epoch, and the top five checkpoints are retained based on validation performance.

\subsection{Fine-tuning Data}
\label{sec:ft_data}

The labelled corpus is assembled from multiple public Indian-language speech corpora\cite{vaani2024,NEURIPS2025_d88714d4,javed2024indicvoices,gangwar2023spring, abhayjeet2025spicor, abhayjeet2025syspin,mile_1,mile_2}
into a single combined manifest, filtered to remove punctuation, digits, and
code-switched or mixed-script transcripts, and lowercased throughout.
Training draws on 17{,}826{,}417 utterances totalling 30{,}565 hours, all of
which fall inside the 0.1--40.0\,s duration window applied at load time;
validation uses 381{,}947 utterances totalling 698 hours, after 2{,}575
utterances (16.8 hours) are discarded by the tighter 0.1--20.0\,s window.

The finetuning corpus spans 65 languages and dialects.
Thirteen exceed 1{,}000 hours---Hindi (3{,}615), Bengali (3{,}193),
Kannada (2{,}142), Marathi (2{,}051), Telugu (2{,}036), English (1{,}709),
Tamil (1{,}441), Assamese (1{,}430), Chhattisgarhi (1{,}405),
Bhojpuri (1{,}247), Malayalam (1{,}105), Magadhi/Magahi (1{,}067) and
Maithili (1{,}048)---and together represent 76\% of training audio, with the
five largest alone accounting for 42.7\%.
A further eleven fall between 100 and 1{,}000 hours: Odia, Punjabi, Bodo,
Gujarati, Manipuri, Dogri, Sanskrit, Nepali, Santali, Sindhi and Konkani.
The remaining 41 are long-tail varieties: five between 10 and 100 hours
(Rajasthani, Chakma, Garo, Nagamese, Mizo), thirteen between 1 and 10 hours
(including Wancho, Garhwali, Marwari, Bajjika, Kokborok, Khortha and Angika),
and 24 below one hour, several represented by fewer than 100 utterances.
This distribution is inherited from the availability of transcribed speech
across Indian languages rather than imposed by sampling, and the resulting
three-order-of-magnitude spread between head and tail is the principal
difficulty the shared 5{,}000-unit vocabulary must absorb.

\subsection{End-to-end Pipeline Summary}
\label{sec:pipeline_summary}

\begin{table}[h]
\centering
\caption{Comparison of the three training stages of SraVaani-1.0.
  \textemdash\ = not applicable.}
\label{tab:summary}
\footnotesize
\setlength{\tabcolsep}{4pt}
\begin{tabular}{llll}
\toprule
\textbf{Item} & \textbf{Stage 1 --- SSL Pretraining} &
\textbf{Stage 2 --- Audio--Image Align.} & \textbf{Stage 3 --- Fine-tuning} \\
\midrule
Model        & FastConformer (SSL) & \makecell[l]{FastConformer\\+ alignment head} & \makecell[l]{FastConformer-Hybrid\\TDT-CTC} \\
Init weights & None (from scratch)       & SSL pretrained checkpoint      & \makecell[l]{Image-Audio aligned  ckpt} \\
Objective    & Contrastive (wav2vec 2.0) & Sigmoid contrastive (SigLIP)   & TDT + CTC ($\lambda=0.3$) \\
Tokeniser    & None                      & None                           & BPE-5{,}000 \\
Dataset      & VAANI (unlabelled)        & VAANI audio--image pairs       & \makecell[l]{Combined multilingual\\corpus (24 sources)} \\
Train hours  & 28{,}418                  & 16{,}580                       & 30{,}565 \\
Val hours    & 1{,}494                       & \textemdash                    & 698 \\
Languages    & \makecell[l]{105\\(165 districts)} & 105          & 65 \\
Precision    & FP32                      & \textemdash                    & bf16-mixed \\
Batch size   & 8{,}192 (eff.)            & 64                             & 192 (eff.) \\
Optimiser    & \makecell[l]{AdamW + Noam\\(2K warmup)} & AdamW (1K warmup) & \makecell[l]{AdamW + Noam\\(10K warmup)} \\
LR           & \makecell[l]{$5\times10^{-3}$ base (Noam)\\peak $3.5\times10^{-6}$} & \makecell[l]{$3\times10^{-4}$ (new)\\$0.05\times$ (encoder)} & \makecell[l]{$10^{-5}$\\(constant in practice)} \\
Duration     & 70 epochs       & 200K steps                     & 50 epochs (max) \\

\bottomrule
\end{tabular}
\end{table}

\section{Evaluation}
\label{sec:evaluation}

\subsection{Evaluation Setup}
\label{sec:exp_setup}

We evaluate \textbf{SraVaani-1.0} , a
multilingual automatic speech recognition (ASR) system built to serve the
linguistically diverse population of the Indian subcontinent.
The model is based on the FastConformer~\cite{rekesh2023fast} architecture
and is trained end-to-end on a large collection of Indic speech data spanning a
wide range of languages, dialects, and acoustic conditions.

\paragraph{Evaluation metric.}
All results are reported in terms of \textbf{Word Error Rate} (\%WER), defined as
\begin{equation}
    \text{WER} = \frac{S + D + I}{N} \times 100,
    \label{eq:wer}
\end{equation}
where $S$, $D$, and $I$ denote the number of substitutions, deletions, and
insertions relative to the reference transcript, and $N$ is the total number of
words in the reference.
Lower WER indicates better performance.

\paragraph{Datasets.}
We benchmark across \textbf{eight publicly available evaluation corpora}
that together cover a broad range of speaking styles, recording conditions,
and language families.
Four consist largely of read speech: \textbf{CommonVoice}~\cite{ardila2020common},
crowd-sourced recordings over 8 Indic languages;
\textbf{FLEURS}~\cite{conneau2023fleurs}, the few-shot benchmark for universal
representations, of which we use 11 Indic splits;
\textbf{IndicTTS}~\cite{indictts2023}, studio-quality
text-to-speech-derived audio across 9 languages; and
\textbf{Kathbath}~\cite{javed2023indicsuperb}, a read-speech benchmark covering
diverse topics in 11 languages.
Three capture more spontaneous or channel-degraded conditions:
\textbf{RESPIN}~\cite{NEURIPS2025_d88714d4}, spontaneous and read speech with regional
dialect variation over 6 languages; \textbf{GramVaani}~\cite{bhanushali2022gram},
community radio speech representing rural and low-resource speaking styles,
evaluated on its Hindi split; and \textbf{MUCS}~\cite{diwan2021multilingual},
the multilingual challenge data, over 6 languages.
Finally, \textbf{Vaani}~\cite{vaani2024} is a large-scale spontaneous image prompted Indic speech corpus in our evaluation, covering many
low-resource and tribal languages absent from every other benchmark.

The Vaani dataset is the primary evaluation resource for
\textbf{44 unique languages} for which no other public benchmark and no
competing ASR system exists (Section~\ref{sec:exp_unique}).

\paragraph{Baselines.}
We compare against three multilingual ASR systems that support at least a
subset of the Indic languages evaluated here.
\textbf{Gemini 3 Flash} (\texttt{gemini-3-flash}) is Google's multimodal
language model with audio transcription capability, evaluated here on Indic
languages~\cite{google2024gemini}.
\textbf{Sarvam Saaras v3} (\texttt{sarvam\_saaras\_v3}) is a commercially
available multilingual Indic ASR model developed by Sarvam AI, supporting the
22 scheduled Indian languages it was built for~\cite{sarvam2024saaras}.
\textbf{IndicConformer-600M-Multilingual}, run with its RNNT decoder, is the open
multilingual Conformer released by AI4Bharat and covers 22 scheduled
languages~\cite{ai4bharat_indicconformer_2025}.
The three span the range of systems a practitioner would realistically
consider: a general-purpose frontier model accessed by API, a commercial
Indic-specialised API, and an open checkpoint that can be self-hosted.

Not all baselines support every language; results are reported only where a
model has an official release for that language, and unsupported
language--dataset pairs are left blank rather than scored.
Table~\ref{tab:model_coverage} summarises the language coverage of each system.

\begin{table}[ht]
\centering
\caption{Language coverage of the evaluated ASR systems.
\checkmark\ = officially supported and evaluated; \textemdash\ = no official support.}
\label{tab:model_coverage}
\small
\setlength{\tabcolsep}{6pt}
\begin{tabular}{lcccc}
\toprule
\textbf{Language category} &
\makecell{\textbf{SraVaani-1.0}\\\textbf{(Ours)}} &
\makecell{\textbf{Gemini}\\\textbf{3 Flash}} &
\makecell{\textbf{Sarvam}\\\textbf{Saaras v3}} &
\makecell{\textbf{Indic}\\\textbf{Conformer}} \\
\midrule
Scheduled Indian languages (20)          & \checkmark & \checkmark & \checkmark & \checkmark \\
English                                  & \checkmark & \checkmark & \checkmark & \textemdash \\
Tribal and regional languages/dialects (44) & \checkmark & \textemdash & \textemdash & \textemdash \\
\bottomrule
\end{tabular}
\end{table}

\paragraph{Implementation details.}
Inference for all models is performed without any test-time adaptation or
fine-tuning.
Text normalisation (lowercasing, punctuation removal) is applied consistently
across all systems before WER computation.
Evaluation scripts follow the standard \texttt{jiwer} convention.

\subsection{Main Results: Multi-Model Languages}
\label{sec:exp_main}

Table~\ref{tab:main_wer} reports WER on the 17 Indic languages for which at
least one baseline model is available for comparison.
Each cell is the unweighted mean over the evaluation datasets in which that
language appears --- between 1 and 8 datasets, given in the \textbf{DS}
column --- so that the comparison is read per language rather than per
individual test split, aggregating 68 underlying language--dataset results.
The best value per language is \textbf{\textcolor{blue}{highlighted}}.

\begin{table}[ht]
\centering
\caption{Word Error Rate (\%, lower is better) per Indic language, averaged
  over the evaluation datasets in which that language appears.  \textbf{DS} is
  the number of datasets contributing to the row and \textbf{Hrs} the total
  duration of the corresponding test audio.  The best value in each row is
  \textbf{\textcolor{blue}{highlighted}}.  $^\dagger$ marks a mean computed
  over fewer than the full set of datasets for that language; such cells are
  excluded from the row comparison.  All four systems report a result for
  every language shown.}
\label{tab:main_wer}
\small
\setlength{\tabcolsep}{5pt}
\begin{tabular}{lrrcccc}
\toprule
\textbf{Language} & \textbf{DS} & \textbf{Hrs} & \makecell{\textbf{Gemini}\\\textbf{3 Flash}} & \makecell{\textbf{Sarvam}\\\textbf{Saaras v3}} & \makecell{\textbf{Indic}\\\textbf{Conformer}} & \makecell{\textbf{SraVaani-1.0}\\\textbf{(ours)}} \\
\midrule
Assamese & 3 & 5.05 & 23.2 & 19.1 & \textbf{\textcolor{blue}{18.5}} & 19.4 \\
Bengali & 6 & 32.33 & 24.7 & 25.0 & 22.7 & \textbf{\textcolor{blue}{19.8}} \\
Gujarati & 5 & 12.55 & 19.1 & 19.5 & \textbf{\textcolor{blue}{19.0}} & \textbf{\textcolor{blue}{19.0}} \\
Hindi & 8 & 51.56 & 14.5 & 14.5 & \textbf{\textcolor{blue}{13.8}} & 14.0 \\
Kannada & 5 & 13.86 & 32.9 & 31.2 & 28.4 & \textbf{\textcolor{blue}{27.4}} \\
Konkani & 1 & 0.18 & 80.9 & 57.3 & \textbf{\textcolor{blue}{53.5}} & 54.7 \\
Maithili & 2 & 5.62 & 50.0 & 44.0 & 46.0 & \textbf{\textcolor{blue}{27.5}} \\
Malayalam & 5 & 9.15 & 35.4 & 33.7 & 28.8 & \textbf{\textcolor{blue}{27.7}} \\
Manipuri & 1 & 0.33 & 103.4 & \textbf{\textcolor{blue}{37.8}} & 39.2 & 41.6 \\
Marathi & 7 & 17.62 & 21.9 & 20.4 & \textbf{\textcolor{blue}{19.4}} & 19.7 \\
Nepali & 1 & 1.15 & 41.4 & 38.0 & 35.9 & \textbf{\textcolor{blue}{21.5}} \\
Odia & 6 & 12.72 & 32.8 & 30.0 & 26.4 & \textbf{\textcolor{blue}{25.7}} \\
Punjabi & 4 & 6.91 & 17.6 & \textbf{\textcolor{blue}{16.2}} & 17.3 & 20.2 \\
Sanskrit & 1 & 3.03 & 36.1 & 57.4 & \textbf{\textcolor{blue}{30.3}} & 36.4 \\
Santali & 1 & 0.12 & 72.0 & 71.0 & 57.9 & \textbf{\textcolor{blue}{57.3}} \\
Tamil & 6 & 30.77 & 35.2$^\dagger$ & 32.1 & 30.1 & \textbf{\textcolor{blue}{26.2}} \\
Telugu & 6 & 17.77 & 28.9 & 28.7 & 26.4 & \textbf{\textcolor{blue}{25.1}} \\
\midrule
\multicolumn{2}{l}{\emph{Mean / total}} & 220.7 & 39.4 & 33.9 & 30.2 & \textbf{\textcolor{blue}{28.4}} \\
\bottomrule
\end{tabular}
\end{table}

\begin{figure}[ht]
\centering
\includegraphics[width=\textwidth]{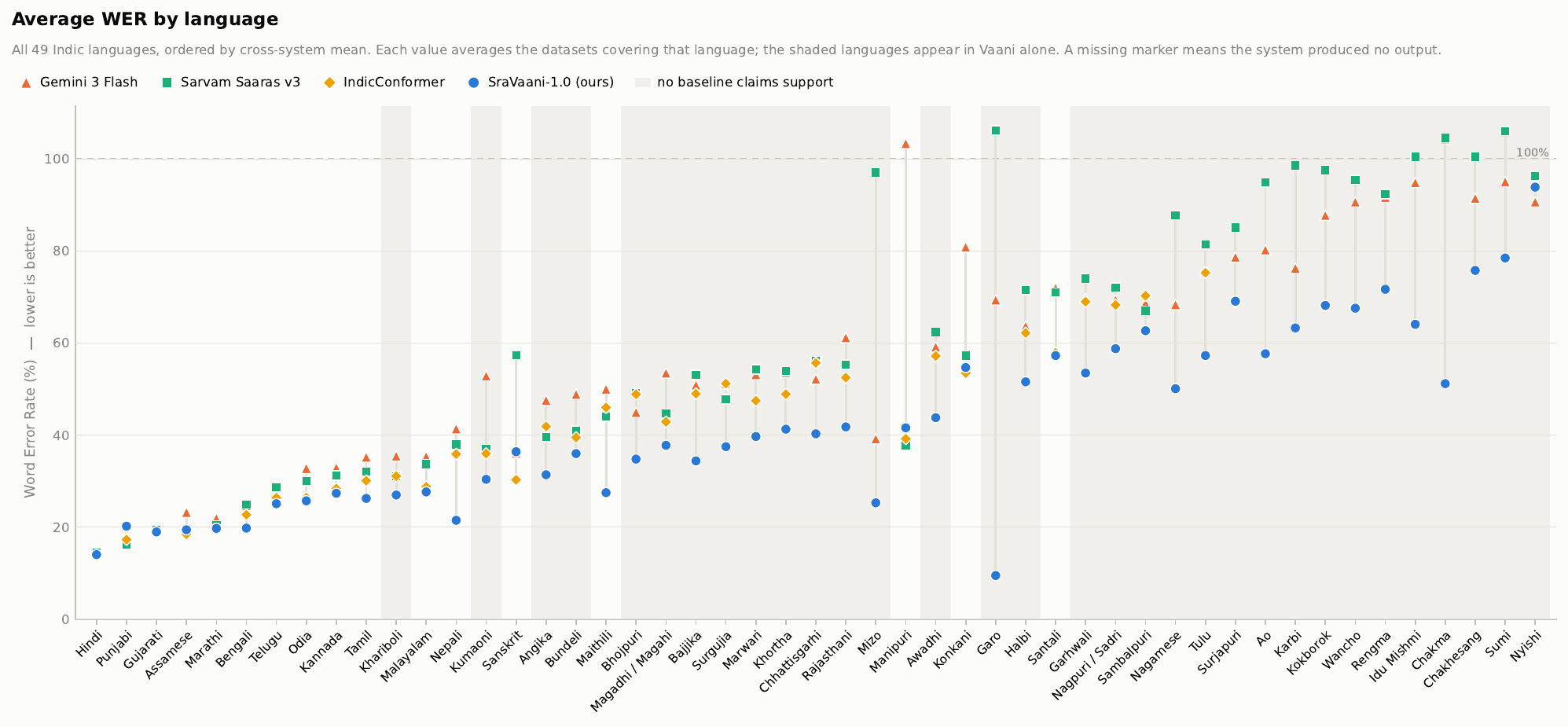}
\caption{Average WER for all 49 Indic languages evaluated, ordered by the mean
  across the four systems.  Each value is the mean over the datasets covering
  that language, so the languages drawn from Table~\ref{tab:main_wer}
  aggregate up to eight corpora while the shaded ones appear in Vaani alone.
  A vertical connector spans each language's best and worst system; a missing
  marker means the system produced no output.  Marker shape as well as colour
  identifies the system, so the figure survives greyscale reproduction.}
\label{fig:wer_all}
\end{figure}
\subsection{Unique Language Coverage}
\label{sec:exp_unique}

A key contribution of SraVaani-1.0 is its ability to transcribe speech
in \textbf{44 languages and dialects} for which no competing ASR system
provides official support.
These span a diverse set of language families including Indo-Aryan,
Dravidian, Tibeto-Burman, Austro-Asiatic, and several tribal and isolate
languages.

Table~\ref{tab:unique_langs} lists each language with its WER on the Vaani
evaluation set, which serves as the only available benchmark for these
languages.
Splits shorter than 0.1\,hours (6 minutes) are omitted: at that size a
handful of utterances moves WER by tens of points, so the figure says more
about the sample than about the model.
This leaves \textbf{32 of the 44} languages, covering 15.0 of the 16.1 test
hours; the model still transcribes the remaining 12, and their results are
available in the release artefacts.

WER values span a wide range.  Garo (9.5\%), Mizo (25.3\%), and
Khariboli (27.0\%) achieve low WER, suggesting strong transfer from related
high-resource languages, and the Indo-Aryan dialects of the Hindi belt ---
Bhojpuri (34.8\%), Bundeli (36.0\%), Magadhi (37.8\%), Chhattisgarhi
(40.3\%) --- cluster in a usable 30--40\% band.
At the other end sit Nyishi (93.9\%), Sumi (78.5\%), and Chakhesang
(75.8\%); these are Tibeto-Burman languages with no high-resource relative in
the training mixture, and they point to the need for dedicated low-resource
strategies.
Across the 32 languages shown the median WER is 50.65\% and the mean 50.2\%,
which should still be read against split size: 23 of the 32 have under 30
minutes of test data.

\begin{table}[ht]
\centering
\caption{Word Error Rate (\%, lower is better) on the Vaani languages and
  dialects that no evaluated baseline claims to support; the Vaani test set is
  the only available benchmark.  \textbf{DS} is the number of datasets
  contributing to the row and \textbf{Hrs} the duration of the test split.
  \emph{None of the three baselines claims support for any language in this
  table.}  IndicConformer requires an explicit language identifier, so it was
  given the supported language whose script matches the target; Gemini 3 Flash
  and Sarvam Saaras v3 were run with no language hint.  The best value in each
  row is \textbf{\textcolor{blue}{highlighted}}; \textemdash\ marks a system
  that produced no output for that language.  The 32 languages with a test
  split of at least 0.1\,hours are shown.  $^\ddagger$ IndicConformer's mean is
  over the 19 languages it produced output for and is therefore not comparable
  with the other three columns.}
\label{tab:unique_langs}
\footnotesize
\setlength{\tabcolsep}{5pt}
\begin{tabular}{lrrcccc}
\toprule
\textbf{Language} & \textbf{DS} & \textbf{Hrs} & \makecell{\textbf{Gemini}\\\textbf{3 Flash}} & \makecell{\textbf{Sarvam}\\\textbf{Saaras v3}} & \makecell{\textbf{Indic}\\\textbf{Conformer}} & \makecell{\textbf{SraVaani-1.0}\\\textbf{(ours)}} \\
\midrule
Angika & 1 & 0.70 & 47.6 & 39.6 & 41.9 & \textbf{\textcolor{blue}{31.4}} \\
Ao & 1 & 0.23 & 80.3 & 94.9 & \textemdash & \textbf{\textcolor{blue}{57.7}} \\
Awadhi & 1 & 0.11 & 59.2 & 62.4 & 57.2 & \textbf{\textcolor{blue}{43.8}} \\
Bajjika & 1 & 0.29 & 50.9 & 53.1 & 49.0 & \textbf{\textcolor{blue}{34.4}} \\
Bhojpuri & 1 & 2.09 & 45.0 & 49.2 & 48.9 & \textbf{\textcolor{blue}{34.8}} \\
Bundeli & 1 & 0.24 & 48.9 & 41.0 & 39.5 & \textbf{\textcolor{blue}{36.0}} \\
Chakhesang & 1 & 0.14 & 91.5 & 100.5 & \textemdash & \textbf{\textcolor{blue}{75.8}} \\
Chakma & 1 & 1.08 & 104.4 & 104.6 & \textemdash & \textbf{\textcolor{blue}{51.2}} \\
Chhattisgarhi & 1 & 1.48 & 52.2 & 56.1 & 55.7 & \textbf{\textcolor{blue}{40.3}} \\
Garhwali & 1 & 0.35 & 69.7 & 74.0 & 69.0 & \textbf{\textcolor{blue}{53.5}} \\
Garo & 1 & 1.07 & 69.4 & 106.2 & \textemdash & \textbf{\textcolor{blue}{9.5}} \\
Halbi & 1 & 0.19 & 63.7 & 71.5 & 62.2 & \textbf{\textcolor{blue}{51.6}} \\
Idu Mishmi & 1 & 0.14 & 94.9 & 100.5 & \textemdash & \textbf{\textcolor{blue}{64.1}} \\
Karbi & 1 & 0.20 & 76.3 & 98.6 & \textemdash & \textbf{\textcolor{blue}{63.3}} \\
Khariboli & 1 & 0.66 & 35.5 & 31.1 & 31.1 & \textbf{\textcolor{blue}{27.0}} \\
Khortha & 1 & 0.41 & 53.6 & 53.9 & 48.9 & \textbf{\textcolor{blue}{41.3}} \\
Kokborok & 1 & 0.64 & 87.8 & 97.6 & \textemdash & \textbf{\textcolor{blue}{68.2}} \\
Kumaoni & 1 & 0.19 & 52.9 & 37.0 & 36.0 & \textbf{\textcolor{blue}{30.4}} \\
Magadhi / Magahi & 1 & 1.05 & 53.5 & 44.7 & 42.9 & \textbf{\textcolor{blue}{37.8}} \\
Marwari & 1 & 0.39 & 53.2 & 54.3 & 47.5 & \textbf{\textcolor{blue}{39.7}} \\
Mizo & 1 & 0.35 & 39.2 & 97.1 & \textemdash & \textbf{\textcolor{blue}{25.3}} \\
Nagamese & 1 & 1.00 & 68.4 & 87.8 & \textemdash & \textbf{\textcolor{blue}{50.1}} \\
Nagpuri / Sadri & 1 & 0.21 & 69.4 & 72.1 & 68.3 & \textbf{\textcolor{blue}{58.8}} \\
Nyishi & 1 & 0.13 & \textbf{\textcolor{blue}{90.7}} & 96.3 & \textemdash & 93.9 \\
Rajasthani & 1 & 0.14 & 61.2 & 55.3 & 52.5 & \textbf{\textcolor{blue}{41.8}} \\
Rengma & 1 & 0.17 & 91.7 & 92.4 & \textemdash & \textbf{\textcolor{blue}{71.7}} \\
Sambalpuri & 1 & 0.22 & 68.7 & 67.0 & 70.3 & \textbf{\textcolor{blue}{62.7}} \\
Sumi & 1 & 0.40 & 95.1 & 106.1 & \textemdash & \textbf{\textcolor{blue}{78.5}} \\
Surgujia & 1 & 0.11 & 51.2 & 47.8 & 51.2 & \textbf{\textcolor{blue}{37.5}} \\
Surjapuri & 1 & 0.14 & 78.7 & 85.1 & \textbf{\textcolor{blue}{69.0}} & 69.1 \\
Tulu & 1 & 0.12 & 75.5 & 81.4 & 75.3 & \textbf{\textcolor{blue}{57.3}} \\
Wancho & 1 & 0.36 & 90.7 & 95.5 & \textemdash & \textbf{\textcolor{blue}{67.6}} \\
\midrule
\multicolumn{2}{l}{\emph{Mean / total}} & 15.00 & 67.8 & 73.6 & 53.5$^\ddagger$ & \textbf{\textcolor{blue}{50.2}} \\
\bottomrule
\end{tabular}
\end{table}

\subsection{Discussion}
\label{sec:exp_discussion}

The results highlight a fundamental \textbf{coverage--accuracy trade-off} in
multilingual Indic ASR.
General-purpose models (Gemini 3 Flash) achieve strong
performance on high-resource languages in clean conditions — best-in-table on
FLEURS Hindi (8.7) and Kathbath Hindi (8.5) — but degrade sharply on
spontaneous speech (53.2\% mean on Vaani) and do not support most Indic
languages.
Specialised models achieve state-of-the-art WER on the languages they
support — Sarvam Saaras v3 on 8 of 11 FLEURS splits, IndicConformer on 5 of 8
CommonVoice splits — at the cost of zero coverage for the remaining languages.

SraVaani-1.0 occupies a unique position: it is the only model
evaluated here that provides \emph{any} transcription capability for
44 languages, and it achieves the best accuracy on 10 of the 17 Indic
languages where comparison is possible (28 of the 68 individual
language--dataset pairs), together with the lowest mean WER (28.4\%) across
those 17 languages.
This breadth is of particular value for downstream applications in
government services, education, and accessibility tools that must serve
speakers of non-scheduled and tribal languages.

\section{Limitations}
\label{sec:conclusion}
The current version of SraVaani has several limitations. First, it does not support Urdu and Kashmiri, as these languages were not included in the finetuning  process although we have used it in the pretraining stage. Second, for code-switched speech, the model generates the entire transcription in the script of the predicted primary language rather than preserving the original scripts of the individual languages, which can reduce readability and linguistic fidelity. Third, the model does not perform text normalization or canonicalization and therefore does not account for valid orthographic or transcription variants that may exist across languages and writing systems. Fourth, the automatic language identification component is not always reliable; in some cases, it misidentifies the spoken language and consequently produces the transcription in an incorrect script, leading to degraded transcription quality. Finally, although the model supports a large number of languages, the amount of supervised fine-tuning data varies substantially across them. For many low-resource languages, only a limited amount of transcribed speech is available, which restricts transcription accuracy and may not yet be sufficient for deployment in production-quality applications. Addressing these limitations will require expanding language coverage, collecting additional supervised data for underrepresented languages, improving multilingual language identification, supporting script-aware code-switched transcription, and incorporating language-specific text normalization techniques.

A further limitation of this study is that both the training and evaluation data are derived from the same underlying dataset. Although the training and test splits are disjoint, they remain in-domain and share similar data collection protocols, recording conditions, and linguistic distributions. As a result, the reported performance may not fully reflect model generalization to truly out-of-domain speech collected under different conditions. Evaluating these models on independent datasets could lead to different conclusions regarding their robustness and cross-domain generalization. This highlights the need for dedicated, publicly available evaluation benchmarks collected independently of training corpora to enable fair, reliable, and reproducible comparisons of multilingual ASR systems across diverse Indian languages.


\begin{ack}
We gratefully acknowledge the support of the AI \& Robotics Technology Park (ARTPARK), Indian Institute of Science (IISc), and Google for enabling this work. We sincerely thank Raghu Dharmaraju (ARTPARK), Prof.\ Bharadwaj Amrutur (ARTPARK, IISc), Dr.\ Partha Talukdar, Dinesh Tiwari, Amritha Kamath, Sukhwinder Singh, and the broader Google leadership team for their invaluable support throughout this effort.
\end{ack}

\small
\bibliography{references}
\normalsize


\appendix


\end{document}